\documentclass{article}

\usepackage{PRIMEarxiv}

\usepackage[utf8]{inputenc} 
\usepackage[T1]{fontenc}    
\usepackage{hyperref}       
\usepackage{url}            
\usepackage{booktabs}       
\usepackage{amsfonts}       
\usepackage{nicefrac}       
\usepackage{microtype}      
\usepackage{lipsum}
\usepackage{fancyhdr}       
\usepackage{graphicx}       
\usepackage{makecell}
\usepackage{orcidlink}

\newcommand{\mycomment}[1]{}

\usepackage[many]{tcolorbox}     
\newtcolorbox{tbox}{
    sharpish corners, 
    boxrule = 0pt,
    toprule = 4.5pt,
    enhanced,
    fuzzy shadow = {0pt}{-2pt}{-0.5pt}{0.5pt}{black!35} 
}
  
\title{ClinConNet: A Blockchain-Based Dynamic Consent Management Platform for Clinical Research
\thanks{\textit{\underline{Citation}}: 
\textbf{M. Naghmouchi and M. Laurent, ClinConNet: A Blockchain-based Dynamic Consent Management Platform for Clinical Research (2026)}} 
}

\author{
  Montassar Naghmouchi \orcidlink{0000-0003-3467-7514}\\
  SAMOVAR, Télécom SudParis, Institut Polytechnique de Paris \\
  \texttt{montassar-bellah\_naghmouchi@telecom-sudparis.eu} 
    \\
    \And
  Maryline Laurent \orcidlink{0000-0002-7256-3721}\\
  SAMOVAR, Télécom SudParis, Institut Polytechnique de Paris \\
  \texttt{maryline.laurent@telecom-sudparis.eu}
}

\begin{document}
\maketitle

\begin{abstract}
Consent is an ethical cornerstone of clinical research and healthcare in general. Although the ethical principles of consent - providing information, ensuring comprehension, and ensuring voluntariness - are well-defined, the technological infrastructure remains outdated. Clinicians are responsible for obtaining informed consent from research subjects or patients, and for managing it before, during, and after clinical trials or care, which is a burden for them. The voluntary nature of participating in clinical research or undergoing medical treatment implies the need for a participant-centric consent management system. However, this is not reflected in most established systems. Not only do most healthcare information systems not follow a user-centric model, but they also create data silos, which significantly reduce the mobility of patient data between different healthcare institutions and impact personalized medicine. Furthermore, consent management tools are outdated. We propose ClinConNet (Clinical Consent Network), a platform that connects researchers and participants based on clinical research projects. ClinConNet is powered by a dynamic consent model based on blockchain technology and take advantage of dynamic consent interfaces, as well as blockchain and Self-Sovereign Identity (SSI) systems. ClinConNet is user-centric and provides important privacy features for patients, such as unlinkability, confidentiality, and ownership of identity data. It is also compatible with the right to be forgotten, as defined in many personal data protection regulations, such as the GDPR. We provide a detailed privacy and security analysis in an adversarial model, as well as a Proof of Concept implementation with detailed performance measures that demonstrate the feasibility of our blockchain-based consent management system with a median end-to-end consent establishment time of under $200 ms$ and a throughput of 250 transactions per second.

\keywords{Clinical Research \and Dynamic Consent Management \and Blockchain \and Self-Sovereign Identity \and Privacy \and Right-to-be-forgotten}
\end{abstract}

\section{Introduction}
Informed consent is a legal obligation that requires physicians and clinicians to obtain the patient' (hereafter referred to as 'participant') consent before proposing any procedure or treatment. In both clinical care and clinical research, informed consent is established via jurisprudence in many countries around the world \cite{Mazur}\cite{Moumjid}\cite{Powers} and is defined by three (3) critical concepts: \textbf{information}, \textbf{comprehension}, and \textbf{voluntariness} \cite{Bazzano}\cite{Nuremberg}\cite{Belmont}\cite{Helsinki}.
Managing consent is a critical part of clinical care and research. Until recently, consent was usually documented in writing, although it could also be given verbally. It is common practice to keep records of consent agreements between physicians and patients, or clinicians and research subjects, in the form of signed documents. In courts, the primary use of informed consent is in retrospective decision-making after injury in the context of clinical care \cite{Mazur}. In clinical research, consent is part of scientific protocols and consent forms are approved by institutional review boards. In both cases, establishing a consent management system is necessary to ensure liability and regulatory compliance.

Consent management systems established in hospitals and clinical research settings often lack integration of digital and dynamic consent tools and systems. Instead of providing patients with a digital interface through which they can manage their consent, these systems require patients to contact a responsible entity via contact forms. In the best cases, institutions offer various digital tools or software that integrate with their Hospital Information System (HIS), Laboratory Information System (LIS) or some other internal digital platforms, whether proprietary or not. The lack of interoperability between these systems suggests a similar lack of interoperability between consent management components. This severely limits the mobility of patient or participant consent data, and by extension health data in general. It also burdens participants with managing fragmented consent across several tools. This is counterproductive to the concept of participant-centric consent management, which is based on the \textbf{voluntary} requirement of informed consent.

In essence, a consent model that enables patients or participants to manage their own consent preferences, free from the limitations of traditional consent management methods, is essential for healthcare and clinical research. Given the patient- and participant-centric nature of healthcare and clinical research,  user-centric identity management and Self-Sovereign Identity (SSI) are the most appropriate models to integrate into this system. However, consent management must also address identity management, as the security and privacy features or limitations of a consent model are inherited from its underlying identity management system. Therefore, a comprehensive dynamic consent model should incorporate identity management, consent, and most importantly, communication tools for clinicians and participants.

\textit{\underline{Our Contributions}}: \\
\textbf{(1) A participant-centric platform:} Thanks to an SSI wallet and a blockchain infrastructure, participants have full sovereignty and control over their identities and identity management. This gives participants total control over their consent, so they don't need to rely on research organizations.  \\
\textbf{(2) Dynamic consent model:} Our consent management operations are executed via smart contracts, giving participants revocable, updatable, non-repudiable consent that they can control and which limits their dependency on research organizations. Moreover, this enables research organizations to automatically register non-repudiable consents and to enable researchers to focus on clinical research rather than managing participant's consent. \\
\textbf{(3) Privacy-by-Design:} Establishing and managing consent offers both participants and research organizations the maximum level of privacy. Our system provides high level of \textbf{unlinkability} and respects the right to be forgotten. \\
\textbf{(4) Full implementation and performance measures:}  To demonstrate the feasibility of our platform, we provide a proof of concept (PoC) for the entire system and all its components. Even for a PoC, our performance measures show promising results: around 250 transactions per second (TPS) for our smart contract functions, and +320 TPS for the most frequent operations. On the SSI wallet, the cryptographic functions related to consent management take an average of under $5 ms$ per operation, with key generation taking around $30 ms$. Full consent can be established on our platform in under $200 ms$. To enable scientific reproducibility, we make our full code base for all components available on GitHub, and provide the performance measurement scripts to facilitate the re-deployment and testing of our platform.

\textit{\underline{Paper Organization}}: 
Section \ref{background} provides background on SSI and blockchain. This section also provides a formal definition of dynamic consent from the state-of-the-art, while 
Section \ref{related-work} explores works proposing dynamic consent models. Our interest lies in the mechanisms and technologies that enable such digital interfaces for managing consent, as well as their use cases.
Section \ref{clinconnet} presents the system model, the functional, security, and privacy requirements, and the system design of ClinConNet (short for \textbf{\textit{Clinical Consent Network}}). We propose a \textbf{web portal} that allows research organizations and volunteers to \textbf{connect} and manage \textbf{identity and consent} during \textbf{clinical} research trials via our \textbf{dynamic consent model} and \textbf{SSI wallet}, both of which are powered by a \textbf{consortium blockchain} infrastructure. 
Section \ref{consent-model} focuses on the dynamic consent management approach, including the interactions between actors to establish and manage consent. 
Section \ref{discussion} provides a full privacy and security analysis. 
Section \ref{impval} describes our implementation of a proof of concept (PoC) and the performance measures obtained, along with their analysis. 
Finally, Section \ref{conclusion} provides a summary of the main results of the paper and an overview of the regulatory, economic and social advantages of our proposal for managing consent in clinical research. It also outlines our future work, which aims to address the limitations of the current proposal and explores the practical applications of our system in the context of the European Digital Identity Wallet and the European Health Data Space.

\section{Background}
\label{background}
This paper makes use of many SSI concepts and standards, as well as certain blockchain features and functionalities. This section provides the background information necessary to understand the technologies and standards employed in the paper.

\subsection{Self-Sovereign Identity (SSI)}
\label{SSI-bg}
\textbf{SSI} is an identity management model very similar to the user-centric model, except it gives users more autonomy, control and ownership over their identifiers and credentials \cite{NIST-Taxonomy}. The SSI model allows users to generate their own identifiers, such as W3C standard \textbf{DIDs} (Decentralized Identifiers) \cite{DID}, and to store their keys, identifiers and identity data on their own devices using an \textbf{SSI wallet application}. An SSI wallet enables users to authenticate and communicate with each other using protocols like \textbf{DID-Auth} \cite{DID-auth}, which allows users to authenticate DIDs by proving that they control the associated private key, and \textbf{DID-Comm} \cite{DID-comm}, which enables two DID owners to exchange data via wallet-to-wallet communication.

A DID is related to a set of cryptographic keys (asymmetric cryptography), a unique identifier, under the control of the user. It is primarily used to identify and authenticate its owner, but it can also be used to sign credentials, tokens, messages, or data, providing a way to authenticate the data and bind them to the DID owner.

Although there are many ways to create an identifier that satisfies the design goals of the DID standard, called \textbf{\textit{did methods}} \footnote{More information on DID methods can be found here \url{https://www.w3.org/TR/did-extensions-methods/}}, we identify two primary types of DID identifiers: public DIDs and private DIDs. A public DID is a discoverable identifier published on a verifiable data registry, such as a blockchain, while a private DID is not discoverable and is only known to a limited number of parties (also referred to as Peer DID or Pairwise DID). SSI requires an infrastructure to operate. This infrastructure publishes public, discoverable identifiers (acting as a Decentralized Public Key Infrastructure DPKI), as well as other data such as revocation information, trusted lists, credential schema and definitions, depending on the design. While most current SSI systems are blockchain-based, blockchain is not a requirement for SSI; the same design goals and principles can be achieved through a decentralized infrastructure \cite{SSI-blockchain}. However, blockchain presents many opportunities for SSI thanks to its features.

\subsection{Blockchain}
\label{blockchain-bg}
\textbf{Blockchain} is a decentralized append-only ledger maintained by different entities through consensus. Data on the blockchain are permanent and immutable, making it perfect for secure long-term record keeping, but less suitable for short-lived, frequently updated data. Blockchains can fulfill various roles within software architecture, providing storage and computing infrastructure with interesting features such as resiliency, fault tolerance, immutability, and decentralized computing. In a trustless infrastructure, blockchain's consensus and data structure (chained blocks) provide a single source of truth that is agreed upon and built upon for future transactions between different parties. These features, alongside the shared governance around which consortium and public blockchains are centered, allow organizations to establish a trustless infrastructure acting as a single source of immutable, non-repudiable truth. This is highly relevant in the context of managing consent between users and service providers.

\subsection{Dynamic Consent}
\textbf{Dynamic consent} is a new approach to consent that promotes the use of interactive digital interfaces to enable more dynamic negotiation, establishment, and management of consent between parties. The two following definitions are provided from the literature:

\begin{tbox}
\textbf{\textit{Definition 1}} \textit{Dynamic consent is a procedure that allows patients and research participants to continuously review and renew their consent to use their private health information.} \cite{Charles}
\end{tbox}
\begin{tbox}
\textbf{\textit{Definition 2}} \textit{Dynamic consent is based on a digital communication interface that connects researchers and participants, placing participants at the heart of decision making.} \cite{Kaye}
\end{tbox}
In this paper, we define dynamic consent as a consent model with a two-way digital communication interface, enabling dynamic interactions and rapid updates and revocations of consents between parties.

\section{Related Works on blockchain-based dynamic-consent management systems}
\label{related-work}

While there is a wide range of research addressing consent management in the health sector and beyond, this section focuses on existing works related to blockchain-based dynamic-consent management systems. We categorize existing solutions into three distinct groups: \textbf{Ledger-Centric} approaches that rely heavily on smart contracts; \textbf{Hybrid Web-Blockchain} approaches that combine web platforms with blockchain storage; and \textbf{SSI-based} approaches that utilize verifiable credentials. 

To rigorously compare these systems, specifically in terms of regulatory compliance, we distinguish between partial and full Right to Be Forgotten (RTBF). We define \textbf{partial RTBF} as systems where consent validity is revoked, but immutable on-chain traces (such as DIDs, schemas, or transaction logs) remain technically linkable to a user. \textbf{Full RTBF} is achieved only when the link between the participant and the on-chain data can be permanently destroyed, rendering the on-chain record an orphan hash with no discoverable association to the identity. Table \ref{tab:related-work-comp-table} compares our proposal with those presented in this section.

\subsection{Ledger-Centric Approaches}
This category relies on smart contracts and on-chain logic to enforce consent rules directly. In \cite{Tith}, the authors propose a \textbf{purpose-based e-consent} model built on Hyperledger Fabric. It utilizes smart contracts and an RBAC (Role-Based Access Control) model to enforce "consent purposes," which are rules defined by patients for different parts of their EHR. Although this proposal enables granular dynamic consent, it combines access control and consent management into a single model. To comply with RTBF, the authors use a unique hash of the patient eID with a random salt to pseudonymize identity. However, this method significantly impacts the performance of data searches.

Similarly, \cite{Merlec} proposes a \textbf{smart contract-based dynamic consent} management system on the Quorum blockchain. It uses a user-centric scheme to provide GDPR-based consent management for any type of personal data. However, relying on smart contracts to manage all aspects of consent poses privacy issues. Because smart contracts are invoked by data subjects and controllers at all times, the metadata of consent agreements—such as participant addresses and operation types—remain visible on the ledger, making full privacy difficult to guarantee.

\subsection{Hybrid web-blockchain approaches}
Hybrid models attempt to balance usability with blockchain immutability. \textbf{Dwarna} \cite{Dwarna}, a web platform provided by the Biobank of Malta, is a highly innovative example of this approach. It uses Hyperledger Fabric to store an immutable record of consents while the web portal manages the user interface. Dwarna is the work closest to ours, yet it still relies on participants contacting the entity or uploading PDF consent forms \cite{dwarna-form}, whereas we use an innovative method involving an SSI wallet to build and sign forms.

In Dwarna, the web portal is the only entity capable of matching a participant's pseudonym to their real identity. This centralization allows Dwarna to support \textbf{full RTBF}: asking the web portal to forget effectively removes the link between the consent record and the participant. However, this conception places a great deal of trust in the portal. Identity management is siloed and centralized by the Biobank, lacking the patient-centric identity management inherent to SSI. This over-reliance on a single organization undermines the decentralized benefits of blockchain. In contrast, our web portal is designed as \textbf{honest-but-curious}, it acts as a proxy but cannot link the consent proof to the user's private DID or to any given research organization.

\subsection{SSI-Based Approaches}
The third category integrates Self-Sovereign Identity. \textbf{MediLinker} \cite{MediLinker} is a platform built on Hyperledger Indy that defines a specific research consent credential to manage clinical research. Dynamic consent is managed via credentials exchanged through a digital wallet, which acts as a digital communication interface similar to our proposal. 

However, MediLinker only supports \textbf{partial RTBF}. While personal data is off-chain, the credential schemas, DIDs, and revocation registries are recorded on-chain. There are no specific mechanisms to handle these on-chain data traces that could technically be linked back to the patient's DID. Furthermore, consent management does not appear to be the central focus of MediLinker, and the research consent credential is neither sufficiently defined nor discussed. Our proposal extends this SSI concept by introducing a specific focus on the consent lifecycle (update/revoke) and ensuring full RTBF through our proxy/anonymizer architecture.

\begin{table*}[ht]
    \centering
    \caption{A comparison of our proposal with other blockchain-based dynamic consent systems.}
    \label{tab:related-work-comp-table}
    \small
    \begin{tabular}{p{2.2cm} p{2.0cm} p{2.2cm} p{2.2cm} p{2.2cm} p{3.0cm}}
        \hline
        \textbf{Proposal} & \textbf{Blockchain} & \textbf{Identity Model} & \textbf{User Interface} & \textbf{Consent Storage} & \textbf{RTBF Support} \\
        \hline
        Dwarna \cite{Dwarna} & Hyperledger Fabric & Centralized (Siloed) & Web Portal & \makecell[l]{On-chain \\ Record} & \makecell[l]{Fully supported \\ (Centralized DB)} \\
        \hline
        \makecell[l]{Purpose-based \\ e-consent \cite{Tith}} & Hyperledger Fabric & User-Centric & Smart Contract & \makecell[l]{On-chain \\ Policies} & \makecell[l]{Fully supported \\ (Off-chain matching)} \\
        \hline
        \makecell[l]{Smart contract-based \\ dynamic consent \cite{Merlec}} & Quorum & User-Centric & Smart Contract & \makecell[l]{On-chain \\ Contracts} & \makecell[l]{Partially Supported \\ (Immutable Logs)} \\
        \hline
        MediLinker \cite{MediLinker} & Hyperledger Indy & SSI (DID) & Digital Wallet & \makecell[l]{Off-chain \\ Credential} & \makecell[l]{Partially Supported \\ (Revocation Only)} \\
        \hline
        \textbf{ClinConNet (ours)} & \textbf{Hyperledger Fabric} & \textbf{SSI (DID)} & \textbf{Wallet +  Web portal} & \textbf{\makecell[l]{On-chain \\ Proof}} & \textbf{\makecell[l]{Fully supported \\ (Permanent Dissociation)}} \\
        \hline
    \end{tabular}
\end{table*}

\section{ClinConNet Design}
\label{clinconnet}
ClinConNet is a digital platform that manages identities and consents for clinical research projects. It connects research organizations with participants based on the projects that these organisations are running and have published on the platform. Participants are \textbf{volunteers} who undergo clinical research trials within a project, generating health data for use by the organization(s). 
Figure \ref{fig:arch} provides an overview of the ClinConNet system.
In this section, we present our system model and components alongside an architectural overview, detailing the functional, security and privacy requirements, as well as the technological choices.

\subsection{System Model}
\label{usecase}

The actors in ClinConNet are defined as follows:

\begin{itemize}
    \item \textbf{Research Organization's SSI wallet}: A research organization creates research projects and their associated \textbf{private consent forms}, which they send to participants through their SSI wallet (edge agent) to get their consent. Organizations are considered to be \textbf{malicious} entities that may attempt to link participants across different projects or reject consent agreements to avoid liability. 

    \item \textbf{Participant's SSI wallet}: Participants are volunteers who interact with the web portal through their SSI wallet (edge agent), and initiate participation requests in research projects from their edge agent. They then receive \textbf{private consent forms} from research organizations' SSI wallet. These forms are used to generate a consent proof in the participant's SSI wallet which is then published on the blockchain via the web portal acting as a \textbf{proxy/anonymizer}. Participants are considered to be \textbf{malicious}; they may attempt to deny that they signed the form or claim that they never consented to a particular version of it.

    \item \textbf{Web portal}: The web portal presents projects and manages the profiles of participants and organizations on the platform via their SSI wallets. It acts as an independent \textbf{proxy/anonymizer}, publishing consent proofs on behalf of participants. It is \textbf{honest-but-curious}, faithfully executing the protocol while potentially attempting to collect information about the research projects the participant is volunteering to or what organizations they are interested in.
     
    \item \textbf{Consortium blockchain}: This decentralized infrastructure is maintained by research organizations, regulatory bodies, and patient advocacy groups. The blockchain is permissioned to promote further governance and legal compliance and to ensure that only authorized actors can interact with the platform. It acts as a Decentralized Public Key Infrastructure (DPKI), enabling the SSI layer and storing and managing consent proofs to facilitate our consent model. It facilitates the identification and authentication of actors on the web portal and amongst themselves. It also stores consent terms and consent proofs (in the form of consent transactions, or "consent TX"), as described in the consent model (cf. Section \ref{consent}). The blockchain provides the necessary business logic (via smart contracts) to manage consent in terms of updates and revocations. Overall, the blockchain platform is considered \textbf{trusted}, as the diversity of its consortium members (including regulators and advocacy groups) ensures reliable consensus, even in the presence of malicious research organizations.

    \item \textbf{Cloud agent}: The cloud agent facilitates wallet-to-wallet communication and acts as a mediator between the research organization's edge agent and the participant's edge agent. It is the cloud part of the SSI wallet, enabling routing of messages between edge agents deployed on actors devices. 
    The cloud agent is \textbf{honest-but-curious}, as it is assumed to route messages between wallets, but it may attempt to perform traffic analysis or metadata correlation to link participants to organizations. 

    \item \textbf{Optional enrollment authority:} 
    The authority verifies the participant's identity and, if successful, enrolls them by issuing a verifiable credential (VC). Participants then use this VC to register with the web portal. This ensures that all registered participants are real individuals, thereby preventing fraud and spam attacks. Adding this optional feature enhances the system's governance and trust.    
 
\end{itemize}
\begin{figure}[ht]
    \centering
    \includegraphics[width=0.8\linewidth]{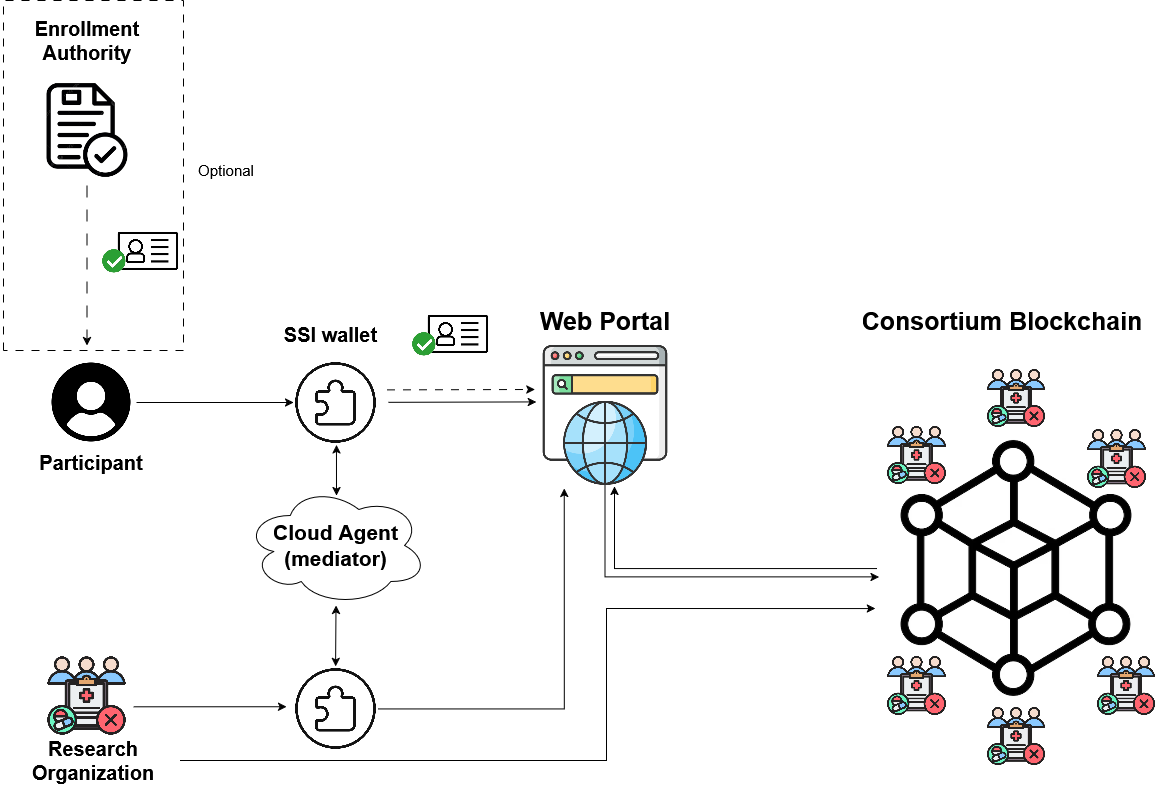}
    \caption{Overview of ClinConNet system model.}
    \label{fig:arch}
\end{figure}

Figure \ref{fig:arch} shows the system actors in interaction. Research organizations (ROs) and participants use their SSI wallet's edge agents to communicate through a mediator which is the cloud agent. Interactions with the web portal need no mediator since the edge agent is a browser extension. Furthermore, unlike participants who interact with the blockchain through the web portal, organizations can also communicate with the blockchain directly since they are authorized to do so.
The dynamic consent model is deployed as a smart contract on the blockchain. This enables both the web portal and research organizations to invoke the smart contract functionalities like reading and writing on the blockchain.
We also represent the optional interaction between the enrollment authority and the participant. This interaction involves verifying the participant's physical identity and associating it with their public DID. It also involves issuing a VC that the web portal verifies during the registration process described in \ref{board-log}.

\subsection{Functional, security and privacy requirements}
\label{design-goals}
We are required to design a system and consent model that allow for \textbf{flexible, continuous and renewable} consent. Moreover, the consent data and consent proof must be \textbf{immutable} (tamper-proof), and \textbf{non-repudiable} (non-refutable), as well as \textbf{transparent}, \textbf{available}, and \textbf{private} (confidential). Dynamic consent also requires \textbf{updatable and revocable} consent that can be updated easily and quickly.
Another important requirement of our model is to ensure the \textbf{unlinkability} of participants, their consents, and their consent proofs. In our consent model, we distinguish between two (2) levels of unlinkability. 
Firstly, we aim to make participants unlinkable across different research projects, which we define as the \textbf{(1) unlinkability of participants across projects}.
Secondly, the consent proof on the blockchain must not be linked to the research organization, the research project, or a participant's public DID, except for the participant themselves. The research organization itself must not be able to link the consent proof to the participant's public identifier. Similarly, the web portal must not link the consent proof to the research organization or the research project. 
This is referred to as the \textbf{(2) unlinkability of consent proofs}.

In addition, we address the \textbf{right to be forgotten (RTBF)}, an obligation under many privacy and data protection laws which blockchain-based systems often struggle to fulfill. 

On the other hand, since we are dealing with dynamic consent, we provide a secure digital interface acting as a two-way digital communication channel, as required by the dynamic consent model (cf. definitions in Section \ref{background}). This interface must allow the interacting entities to establish a \textbf{mutually authenticated} and \textbf{confidential} (end-to-end encrypted) communication. This secure communication channel will be used to exchange and establish consent between participants and research organizations. In line with the \textbf{patient-centric} principle found in the dynamic consent definitions, we also want this secure digital interface to be \textbf{patient-centric} and \textbf{decentralized}, ensuring that the participants are at the heart of our system and in control of their identity and consent data.

\subsection{The edge agent}
\label{edgeagent}
The edge agent, which is one component of the SSI wallet, is hosted on the user's device as a browser extension. The user in question may be either the research organization or the participant.

The edge agent has two main types of DID identifiers:
\begin{itemize}
    \item \textbf{Public DIDs}, which are published on the blockchain. Participants use this public DID to authenticate with the web portal and to manage their consent. The research organization (RO) has a single public DID, used to identify and authenticate the organization within the system (with participants, other organizations and the web portal). Participants have a single public DID for authenticating with the web portal. According to the legal requirements relating to consent types, the optional enrollment authority may link the public DID to the participant's true identity using the issued VC.  
    \item \textbf{Private DIDs}, which are independently created by participants for each research project in which they wish to participate. Participants must identify and authenticate themselves to the research organization using their chosen private DID. Participants are only known to the organization thanks to their private identifier, ensuring the public DID of a participant is never disclosed to a research organization.
\end{itemize}

Thanks to DIDs and associated private keys generated on their edge agents, participants and organizations can perform digital cryptographic operations such as signing and verifying consent forms. The edge agent also enables participants to generate consent proofs to be published on the blockchain (cf. Section \ref{consent}). To improve privacy, the participant's edge agent must ensure that the web portal always lists a sufficient number of projects.

Edge agents are essential for establishing secure communications. First, they provide an end-to-end authenticated and encrypted channel between the participant and the organization via the cloud agent. Second, they provide an authenticated channel to the web portal. To improve privacy, it is assumed that the participant's edge agent interacts with the system using Tor technology. Finally, to prevent the web portal from linking participant activity, the edge agents ensure that there is a sufficient number of projects listed by the web portal.

The edge agents of the participant and organization are assumed to store all consent-related data (private consent forms, consent proofs and TX references of consent proofs) locally on the edge device. 

\subsection{The cloud agent}
\label{cloudagent}
The cloud agent is the second component of the SSI wallet, and acts as a mediator, facilitating end-to-end communication between the participant's and the research organization's edge agents. 
It implements SSI protocols like DID-Auth and DID-Comm (cf. Section \ref{SSI-bg}) and establishes an authenticated and encrypted communication channel between the requesting participant (\textbf{private DID}) and the research organization (\textbf{public DID}). 
The cloud agent must also provide DID resolution to participants, enabling them to authenticate research organizations.

\subsection{The web portal}
\label{webportal}
The web portal acts as an application server. It publishes ongoing research projects and provides a search engine that participants can use to search for projects according to certain criteria. 
It also acts as a proxy/anonymizer for participants. It publishes public DIDs and consent proofs on behalf of participants, thus preventing curious entities from linking a public DID or consent proof to a specific participant, project, or organization. The objective is to ensure herd privacy. 

The web portal has a blockchain identity and can authenticate itself to the permissioned blockchain. It can also interact with the blockchain, and invoke smart contracts that manage identity and consent, all within certain permissions.

The web portal must also store the association between participants' public DIDs and all the proofs of consent that they request to be published on the blockchain, along with the TX reference on the blockchain and the current status, which is either ‘valid’ or ‘revoked’. Thanks to the local matching records kept by the web portal, the web portal can provide the concerned participant with a full history of their consents. It also provides direct functions that enable participants to revoke or update their consent proofs. 

To support RTBF (cf. Section \ref{design-goals}) in the event that consent is requested for revocation, the web portal only needs to remove the consent proof from the list associated with a public DID. This permanently dissociates the proof from the participant.

\subsection{The consortium blockchain}
\label{blockchain}

The blockchain supports the identification and authentication functions of SSI, as well as consent proof management. 
A smart contract is deployed on the blockchain to publish, update, revoke and query consent proofs and consent terms. The blockchain provides accountability, because every output of the smart contract is recorded on the blockchain once consensus has been reached. Furthermore, the publication, updating and revocation of consent proofs only take effect once they have been written on the blockchain, meaning they are all validated by the consortium. This further enhances the security and trustworthiness of the consent model.

In our threat model, in which research organizations may act maliciously, the blockchain consortium includes not only research organizations, but also regulatory bodies, hospitals, and patient advocacy groups. This allows us to consider the blockchain as a trusted entity.

The smart contract is deployed on the blockchain, providing the necessary functions, such as PUBLISH\_CONSENT\_TERMS, which a research organization can use to publish the public consent terms of a research project, and PUBLISH\_CONSENT\_PROOF, which the web portal can use to publish consent proofs generated by participants. The smart contract also provides the QUERY\_CONSENT\_PROOF function, which securely fetches consent proofs and terms from the blockchain. Moreover, we have implemented a REVOKE\_CONSENT function that can be used to revoke or update consent. Our update logic simply involves revoking the old consent proof and publishing an updated one using the supersede-and-replace policy. This update and revocation logic may differ from one blockchain platform to another, but it can be implemented on all platforms. In Hyperledger Fabric for example, the consent proof is an object with different properties (timestamp, published ID, proof hash, etc), and, most importantly, it can have a \textbf{state} like \{'valid' or 'revoked'\}.
The revocation function changes the state of a consent proof from 'valid' to 'revoked', thereby revoking the consent proof on the blockchain.

\subsection{Technological orientations and relationships}
To better understand how the solution was designed, Table \ref{tab:2} provides a mapping between the requirements set out in Section \ref{design-goals} and the technological design decisions made.

\begin{figure}[ht]
    \centering
    \includegraphics[width=0.8\linewidth]{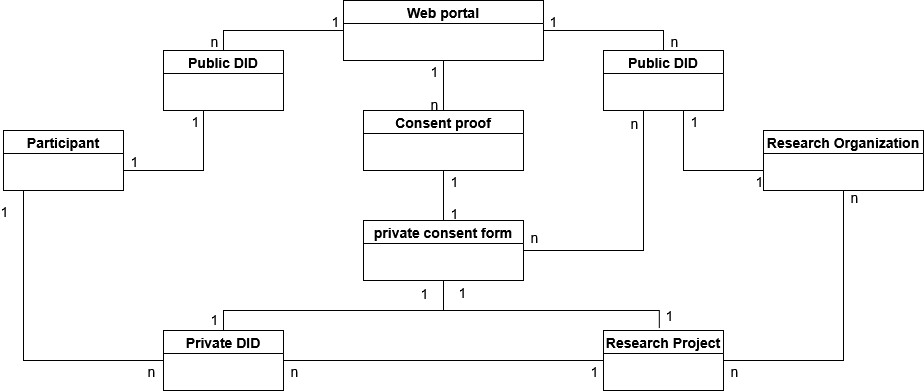}
    \caption{One-to-many relationships between actors, identifiers, consents and projects.}
    \label{fig:relation}
\end{figure}

Figure \ref{fig:relation} illustrates the one-to-many relationships between research organizations, participants, identifiers, projects, consent forms and consent proofs, and the web portal, and is useful for identifying potential linkability issues. 

\begin{table}[ht]
    \centering
    \begin{tabular}{c c}
        Requirements & Design decision \\
        \hline
        Patient-centric & SSI as an identity management model \\ & and communication interface \\
        \hline
        Decentralized system & Blockchain as an infrastructure for both SSI \\ & and consent model \\
        \hline
        Flexible, continuous and renewable consent & Dynamic consent model \\
        \hline
        Updatable and revocable consent & Smart contract automation for updating 
        \\ & and revoking consent \\
        \hline
        Tamper proof consent data & Create a consent proof on the blockchain to verify \\ & the integrity of the consent data \\
        \hline
        Private (confidential) consent data & Consent forms exchanged on a secure \\ & private communication interface \\
        \hline
        Private (confidential) consent proof & Consent proof as a hash of encrypted consent data, \\ & published via a proxy on the blockchain for herd privacy \\
        \hline
        Immutable, transparent and available & On-chain consent proofs (on the blockchain) \\ consent proof & \\
        \hline
        Non-repudiation of consent & Consent is signed by both the participant and the research 
        \\ and consent proof &   organization and consent proof signed by the participant and 
        \\ &  approved by the portal through its publication on blockchain \\
        
        \hline
        Unlinkability of participants (1) & Participants generate different private DIDs \\ & for different research organizations \\
        \hline
        Unlinkability of consent  & Consent is only linked to a private DID \\ and consent proofs (2) & and consent proofs are published via a proxy/anonymizer \\
        \hline
        Mutually authenticated communication & Establish authenticated SSI channels using DID-Auth \\
        \hline
        Confidential communication & End-to-end encrypted messages using DID-Comm \\
        \hline 
        RTBF & On-chain proofs can be permanently dissociated from  \\ & the participant  \\
        \hline
    \end{tabular}
    \caption{Mapping system requirements to design decisions.}
    \label{tab:2}
\end{table}

Both participants and research organizations have a single public DID. Participants generate independent private DIDs to interact with different organizations, establish multiple identities, and manage consent within research projects.
The research organization creates a private consent form for each project, binding it to their public DID and the private DID of the participant. This consent form is then signed using keys associated with both DIDs. The participant is responsible for generating a consent proof from the private consent form.
The participants' public DID is only used to establish an authenticated channel with the web portal, which they then use to instruct the web portal (anonymizer) to publish the consent proof on their behalf. 

The web portal maintains a local match between the consent proof and the public DID. We fulfill the RTBF by removing this local match. As there is no remaining link between the public DID of the participant and the consent or consent proof, the participant is effectively permanently dissociated from the on-chain consent proof.

\section{Dynamic consent management }
\label{consent-model}
This section explains how consent is managed on our platform through a combination of SSI operations and smart contract functions.

\subsection{SSI connection between actors}
\label{board-log}
\begin{figure}[ht]
    \centering
    \includegraphics[width=1\linewidth]{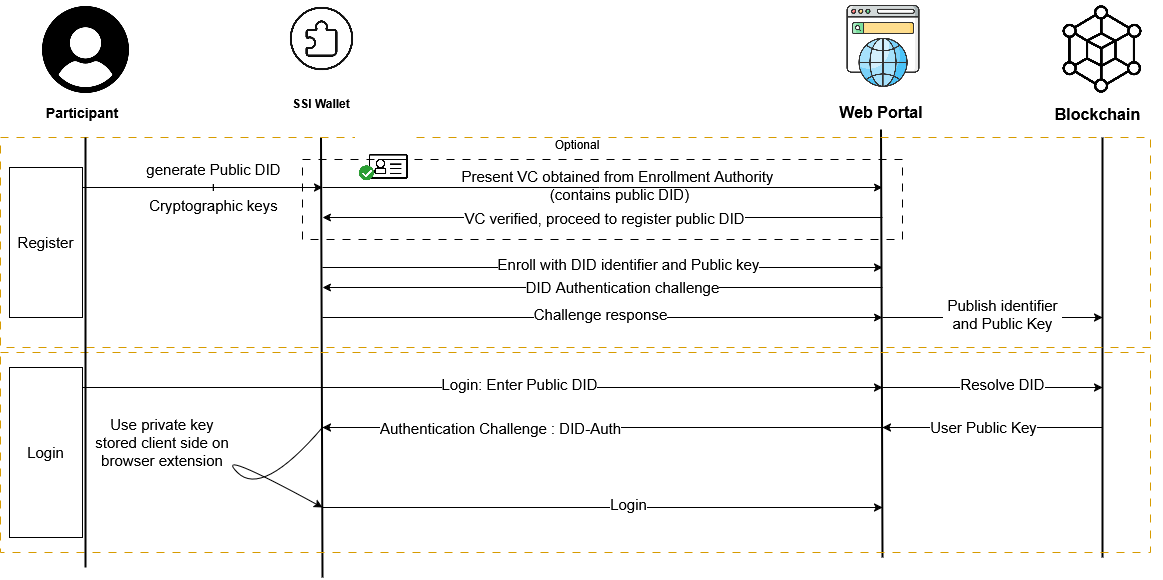}
    \caption{Sequence diagram - Participant onboarding and login on ClinConNet.}
    \label{fig:reglog}
\end{figure}

As shown in Figure \ref{fig:reglog}, participants do not need a password to register or log in to the web portal; instead, they use public DIDs and their associated keys. Participants only have to prove that they know the private key associated with the public DID using a simple challenge-response protocol. Onboarding can be conditional upon proving ownership of a VC issued by the enrollment authority. This enables the web portal to register only relevant participants. 

Logging in to the web portal is done via DID-Auth. The portal generates a public DID-Auth challenge to which the participant's wallet can respond by providing an authentication token back to the web portal. This token is signed by the participant's private key associated with its public DID, which effectively proves his identity. Once registration is complete, the web portal creates a profile for each participant, enabling them to easily browse and select clinical research projects to volunteer in.

\begin{figure}[ht]
    \centering
    \includegraphics[width=1\linewidth]{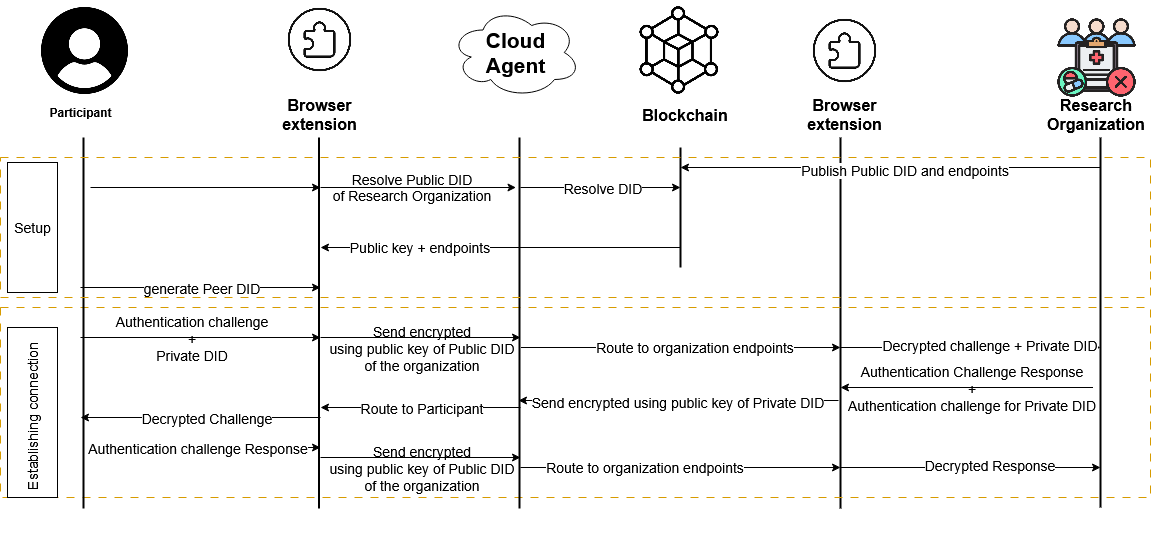}
    \caption{Sequence diagram - Establishing a mutually authenticated connection between the wallets of a participant and a research organization.}
    \label{fig:connection}
\end{figure}

Once a project has been selected and the public DID of the research organization has been retrieved, participants must set up a mutually authenticated channel in order to request the research organization to participate.
As illustrated in Figure \ref{fig:connection}, during the setup phase, the participants resolves the organization's public DID to obtain its endpoints and public key. They then generate a private DID for the project, and establish an authenticated connection with the organization (via their endpoints) through the mediator (cloud agent) using DID-Auth. Participants authenticate the organizations using the public DID resolved earlier, and the organizations authenticate participants using their private DID. 

\subsection{Consent Establishment}
\label{consent}

\begin{figure}[ht]
    \centering
    \includegraphics[width=1\linewidth]{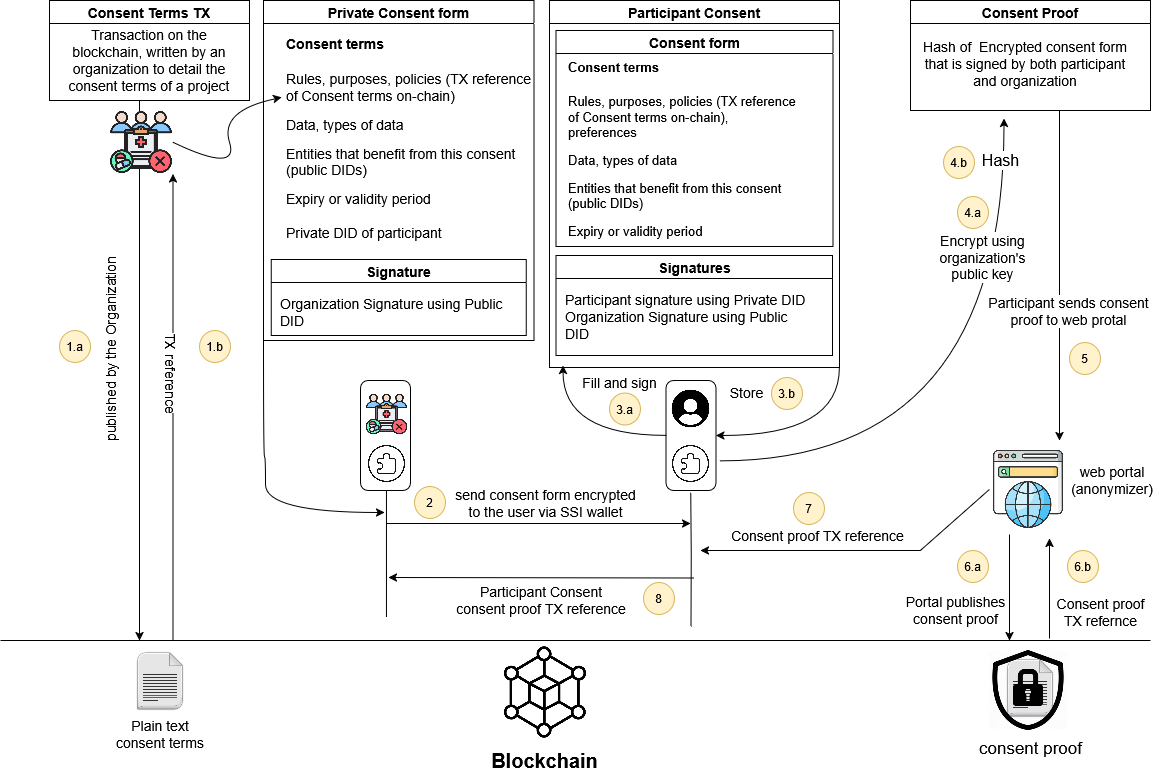}
    \caption{Data flows and artifacts - the end-to-end process of formulating consent.}
    \label{fig:bdc2}
\end{figure}

As illustrated in Figure \ref{fig:bdc2}, the consent model starts with the organization publishing the general consent terms of a research project on the blockchain (step 1.a). The blockchain transaction containing the \textbf{Consent terms TX} of a given project is specified in the project published on the web portal and in any private consent form (step 1.b). This enables participants, supervisory boards and institutions to transparently monitor and audit the project. 
The participant sends a participation request to the research organization after establishing a connection, as described in Section \ref{board-log} and Figure \ref{fig:connection}. The organization then creates a signed \textbf{private consent form} for the participant, mentioning their private DID and encrypting it for them using the public key associated with that private DID (as a DID-Comm encrypted message). This private consent form is signed by the research organization and must be completed and signed by the participant.

Upon receiving the private form (step 2), which is transmitted confidentially between wallets, the participant (1) verifies the organization's signature; (2) reads and completes the private consent form; (3) signs the form with the private key associated with their private DID (step 3.a); (4) stores a local copy (step 3.b); (5) encrypts the completed signed consent form with the organization's public key (step 4.a); (6) hashes the encrypted completed signed consent form to create the \textbf{consent proof} (step 4.b); (7) requests the web portal (anonymizer) to publish the consent proof on the blockchain on their behalf (steps 5 and 6), receiving back a \textbf{consent TX} reference (step 7); and finally (8) sends the encrypted form, the consent proof and the consent TX reference to the organization (step 8). The organization can finalize this process by checking that the hash of the encrypted form matches the consent proof, and that the consent proof's status on the blockchain is set to \textbf{'valid'}.

\subsection{Update and revocation of consent}
\label{consent-management}

\begin{figure}[ht]
    \centering
    \includegraphics[width=1\linewidth]{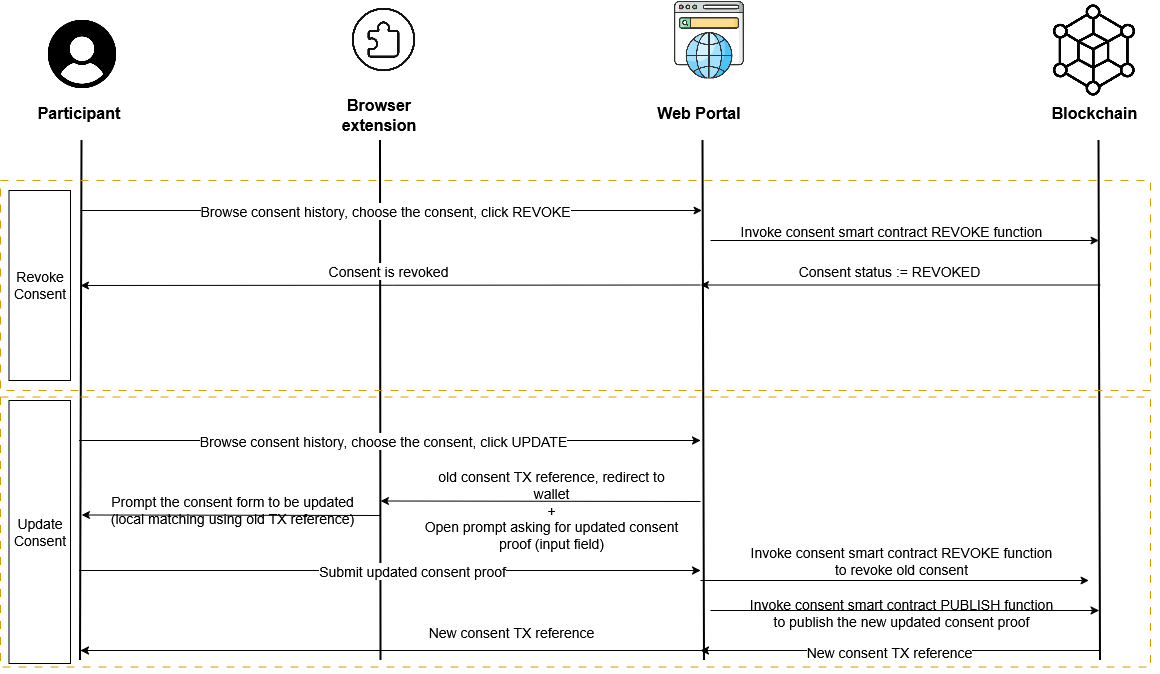}
    \caption{Sequence diagram - Revocation and update of consent.}
    \label{fig:consent-update}
\end{figure}

As shown in Figure \ref{fig:consent-update}, participants can access a history table of all of their consents on the web portal thanks to local matching. They can also locally match this table with the consent data stored in their SSI wallet, using the consent TX as the matching key. Eventually, the participant's SSI wallet contains a full consent history, including the blockchain proof TX reference, timestamps, consent data, associated private DIDs, the organization's public DID, project consent terms and description.
By default, the 'state' (status) property of consent proofs published on the blockchain is set to 'valid' upon publication. The consent smart contract uses the revoke function to change the consent status from 'valid' to 'revoked', which indicates that consent has been withdrawn. Consent updates follow a supersede-and-replace policy, whereby an update  effectively invokes the revoke function on the previous consent and publishes a new consent proof.

The status of the consent proof on the blockchain provides irrefutable proof of updates and revocations. The fact that this runs on a blockchain enforced by a consortium limits the ability of research organizations to ignore such requests or delay their execution.

\section{Privacy and Security Analysis}
\label{discussion}
This section analyzes the privacy and security properties of ClinConNet (cf. Section \ref{design-goals}), under an adversarial model where the web portal and the cloud agent are considered honest-but-curious, while participants and research organizations are malicious.
The analysis focuses to verify that the platform does not disclose any information resulting from digital interactions and stored information. This means that the analysis does not take into account personal data that participants may disclose via the consent form or during clinical trials. 
It should be noted that there are indeed cases where this non-disclosure is guaranteed, for example with anonymous or pseudonymous consents for clinical projects that do not require the patient's identity to be known (e.g, collection of vital signs during sleep from participants wearing a smart device).

\textbf{Unlinkability of the participant across projects}: 

The three following attack scenarios are analyzed: 
\begin{itemize}
    \item Colluding organizations: 
     In order to link one participant across projects, the organizations hosting the projects might collude in an attempt to match that participant's different private DIDs or link them or their consent proof to a given public DID.
    Neither attempt can succeed, since (1) a participant generates a unique private DID for every project they volunteer for, (2) private DIDs are generated independently of one another, (3) there is no relationship between the public DID and the consent form or the on-chain consent proof. The only link is a local matching stored by the web portal.
    \item Web portal: The web portal is unaware of the content of the consent proofs it publishes. What it can attempt to do is match the research projects sought by the participant with a proof. However, it will not be able to link the proof to a specific project because the edge agent always checks that the web portal lists a sufficient number of projects (cf. Section \ref{edgeagent}).
    \item Cloud agent: The cloud agent does not have access to the content of encrypted traffic between two edge agents. However, it may attempt to link a participant to the organization's edge agent by collecting the network metadata (e.g, IP addresses) of the participant's edge agent. As indicated in \ref{edgeagent}, Tor technology is used to decouple the network identity from the application identity (public DID) of the participant's edge agent, preventing the attack. 
    
\end{itemize}

\textbf{Unlinkability of consent proofs}: 

The discussion of the previous property leads to the conclusion that the research organization hosting the project cannot link a consent proof to a public DID of a participant, and web portal to a research project or the public DID of an organization.

Consider any external entity, i.e. one that is neither the web portal, the participant nor the hosting organization. This entity can only observe the consent proof on the blockchain in the form of an opaque hash value. Without the original consent form and the research organization's private key, it is impossible to link the consent proof to an organization, participant or project. Therefore the property is provided.

\textbf{Confidentiality of consent and consent proof}: \\
 We refer to the following two attack scenarios: 
\begin{itemize}
    \item The cloud agent attempts to read the consent form during the consent agreement phase between the edge agents of the participant and the research organization. This attack is not possible because the content of the traffic is end-to-end encrypted using DID-Comm message envelopes.
    \item Any entity external to the system can read the consent proof on the blockchain, but is unable to extract useful information from it, as already mentioned in the previous property discussion. 
\end{itemize}

\textbf{Integrity and non-repudiation of consent and consent proof}: \\
Integrity and non repudiation are closely related. These are ensured by the immutability feature of the blockchain, and the manner in which consent data is produced. 
There are two attack scenarios: 
\begin{itemize}
    \item Participants may attempt to modify their signature or claim they did not consent to a specific version of a consent form. This fails for the following reasons: (1) the participant must sign the consent form using the private key associated to their private DID; (2) the organization can  cryptographically check the validity of the signature because they know the public key corresponding to the private DID; (3) the organization can check that the on-chain consent proof corresponds to the agreed consent; (4) the participant authenticates with their public DID on the web portal for registering the consent proof transaction. This strongly links the consent proof to a specific participant; (5) finally the web portal can reveal the public DID of the participant which uniquely identifies them in the system. If the optional enrollment authority is operational and a non-repudiation investigation is conducted, it is even possible to reveal the real identity behind the public DID. 
    
    If a participant modifies their consent proof before publishing it on the blockchain, the organization considers the consent invalid due to step (3). If they deny having given their consent, but it is registered on the blockchain, the non-repudiation investigation in step (5) will reveal their true identity (or at least their public DID) and prove they are lying.
    
    \item The organization attempts to repudiate a consent agreement in order to avoid liability. This fails for the following reasons: (1) the organization signed the consent form using their public DID's associated private key, (2) the participant kept the consent form signed by the organization in their wallet; (3) the consent proof was published on the blockchain. The participant can re-generate the proof from their wallet, comparing it to the published value. Therefore the participant is able to prove that the organization agreed on a specific consent and is lying. 
    \end{itemize}

\textbf{Availability of the full dynamic consent control}: 
Consent data remains available to participants and organizations in the SSI wallets off-chain. Although additional wallet recovery mechanisms could enhance security, these are beyond the scope of our contribution.
The availability of on-chain data like public DIDs, public consent terms and consent proofs is guaranteed by blockchain properties.
The dynamic control over the consent model is available for the following reasons: (1) Smart contracts that enforce control logic (Update/Revoke), are available due to the decentralized nature of blockchain technology; (2) the web portal acts as a proxy to trigger these contracts, and as an honest but curious entity that faithfully executes participant's requests to revoke or update consent.

\textbf{RTBF (Right To Be Forgotten)}: 
The web portal is the only entity capable of implementing the RTBF. The web portal is honest but curious. Therefore, upon request of a relevant authenticated participant, it will delete the corresponding entry in its database for consent proof to be forgotten, even in the case of malicious organizations. Moving the consent proof status to deletion indicates strong RTBF claims that the organization - being malicious - can ignore but can never deny. This provides strong legal liability.

\section{Implementation and performance analysis}
\label{impval}
This section presents the proof of concept (PoC), and performance results, demonstrating the feasibility of our ClinConNet platform. We specify the hardware and software specifications for the implemented PoC. We also present the methodology that explains how we tested our system, and provide the results and their analysis. The ClinConNet code base is available on GitHub \ref{code-base} alongside the performance measures, scripts and results, ensuring the reproducibility of our work.

\subsection{Testbed and PoC description}
\label{imp}
\begin{figure}[ht]
    \centering
    \includegraphics[width=0.8\linewidth]{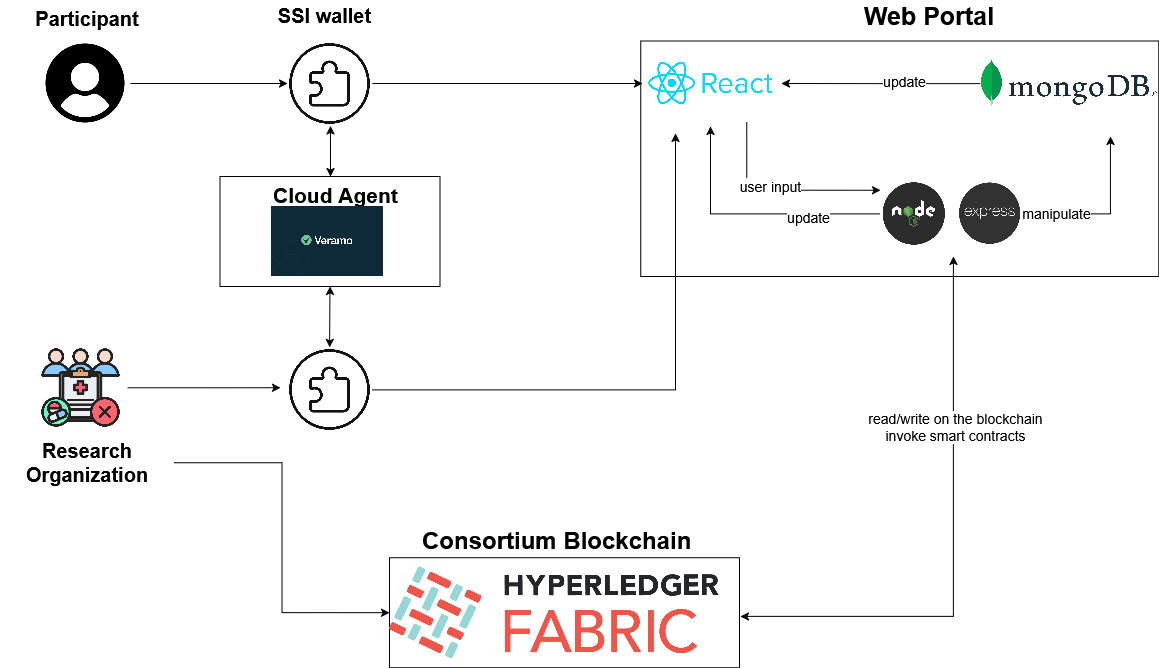}
    \caption{An overview of the implemented architecture and the chosen technologies.}
    \label{fig:softmatcharch}
\end{figure}

Figure \ref{fig:softmatcharch} illustrates how the PoC technologies map onto our architectural components in Figure \ref{fig:arch}, and Table \ref{tab:software} lists the software versions that were used. 
We use \textbf{Hyperledger Fabric} as a permissioned blockchain and the \textbf{MERN} stack (MongoDB, Express, React and Node.js) stack to implement the web portal. As for the SSI wallet, we use Veramo \cite{Veramo-github} wallet SDK.
Hyperledger Fabric uses a Membership Service Provider (MSP) for permission management via Certificate Authority (CA) servers and X.509 certificates. The MSP allows all organizations, the cloud agent and the web portal to securely interact with the blockchain. We use the test network of Hyperledger Fabric, which consists of two peer nodes and one orderer node.

The PoC is implemented on a single machine, a \textbf{Dell Precision 5480}, with the \textbf{13th Gen Intel CPU core i9-13900Hx20} running \textbf{20 Cores} at \textbf{1.39 Ghz}. The machine has \textbf{32 GB of RAM}.
The \textbf{OS} is a \textbf{Ubuntu 24.04.1 LTS} of \textbf{64-bit}.

\begin{table}[ht]
    \centering
    \begin{tabular}{p{100pt}p{100pt}p{100pt}}
         Component & Technology & Version \\
         \hline
         Blockchain & Hyperledger Fabric & v2.2 \\
         \hline
         Web portal  & MongoDB & v7.0.17 \\
          & React & v19.0.0 \\
         &  Express & v4.21.2 \\
         & Node.js & v18.19.1        \\
         \hline
         SSI Wallet & Veramo & v6.0.0 \\
         \hline
    \end{tabular}
    \caption{Software versions used in the ClinConNet proof of concept.}
    \label{tab:software}
\end{table}

\subsection{Methodology}
\label{methods}
The performance evaluation is designed to put the two main technological components of ClinConNet under stress: the browser-based SSI wallet (edge agent), which runs on the Chromium browser, and the smart contracts that run on the blockchain to automate the dynamic consent model. 
The web portal and cloud agent were deliberately excluded because they have no cryptographic burden and are standard web servers with no novel architectural bottlenecks. The SSI wallet of the research organization was also excluded since the SSI operations run in a stable, high-performance server environment using the same cryptographic primitives as the participant's wallet.

We describe the methodology for the three performance measures that we carried out; the results are documented in Sections \ref{ssi-perfs}, \ref{blockchain-perfs} and \ref{perf-consent}.

\textbf{Key operations on the participant's SSI wallet: }   
We evaluate the impact of identity and cryptographic operations on the participants' user experience. We measure the execution time of our five main edge agent functions: generating a DID using the did:key method, verifying a digital signature found on a consent form, signing the consent form, generating a consent proof, and finally performing DID authentication requests, which includes generating a JWT (JSON Web Token) for authentication using DID-Auth. 
We use \textbf{puppeteer}\footnote{\url{https://pptr.dev/}} to evaluate the performance of the SSI wallet and automate the tests. This tool enables precise and non-intrusive timing of JavaScript execution within the browser. We test these functions by repeating each one 100 times (10 iterations in 10 runs), and repeating the whole process three (3) times to ensure the results were not volatile. This repetitive approach helps to capture the execution of the background workers within the application, and ensures that the data reflects sustained operational performance rather than transient peaks.
The GitHub repository containing the test scripts and their results is provided in \ref{code-base}.

\textbf{Hyperledger Fabric smart contracts:} We evaluate the performance of the three main functions of the blockchain smart contract which are: publishing a consent proof, revoking consent, and querying a consent proof. 
We use \textbf{Hyperledger Caliper}\footnote{\url{https://www.lfdecentralizedtrust.org/projects/caliper}}, a standard benchmarking tool for private blockchain instances, to automate the testing of smart contract functions on the Hyperledger Fabric blockchain. Caliper enables us to test the performance and scalability of the blockchain platform and the dynamic consent smart contracts.
We start by testing our blockchain with a fixed send rate of 50 transactions per second (TPS). 
We test write transactions like PUBLISH\_CONSENT\_PROOF and REVOKE\_CONSENT and read transactions like QUERY\_CONSENT\_PROOF. 
As write operations are more costly than read operations, but read operations are more frequently requested in our clinical trial use case, we double the number of read transactions to put stress on our blockchain system and smart contracts. 
Similar to the SSI wallet tests, the testing process is repeated three times independently. In the first two runs, we send a fixed rate of 50 TPS for a total of 500 transactions (TX) for the publish and revoke functions (each), and for read transactions a fixed rate of 100 TPS for a total of 1,000 read transactions. For the third and final run, the extreme evaluation, we send at a rate of 500 TPS for a total of 5,000 transactions for the publish and revoke functions, and at a rate of 1,000 TPS for a total of 10,000 transactions for the query function. This is ten times more intense than the previous two runs. This enables us to reach a statistical steady state, allowing us to observe how our system behaves under pressure and determine whether services degrade due to failed transactions. It also enables us to test the scalability of our consent model by increasing the load tenfold. Benchmarking studies targeting clinical trial management systems (CTMS) on blockchain, like \cite{benchmark1}, demonstrate that 50-300 TPS is sufficient for handling millions of transactions in a global research setting.  Our send rate of 50–500 TPS is intended to verify whether we are within the competitive range of related works or not.

\textbf{Consent model end-to-end lifecycle:} In order to assess whether the entire consent management process is compatible with a positive user experience, we evaluate the time taken for the entire lifecycle of consent management within our system. If the full process is compatible with a good user experience, we can conclude that every single operation is also compatible. Consequently, we sum the key internal execution times (their median values) of the different operations involved in consent management. This sum is called an end-to-end lifecycle median time. This provides a performance indicator that excludes non-controllable factors like network latency and the human factor, allowing us to isolate and prove our core logic and architecture.

\subsection{Performance analysis for key operations on the participant's SSI wallet}
\label{ssi-perfs}
\begin{figure}[ht]
    \centering
    \includegraphics[width=1\linewidth]{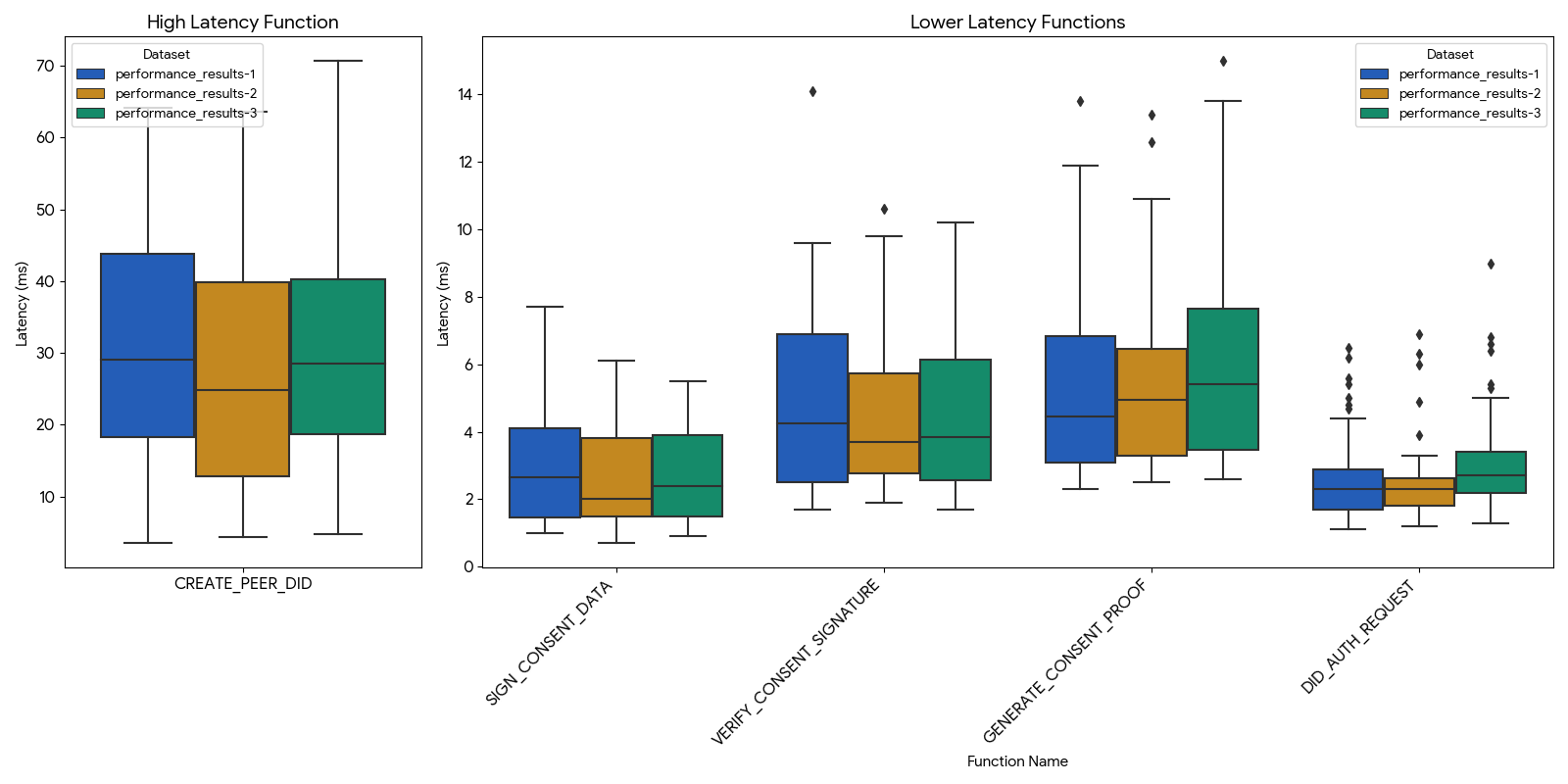}
    \caption{Latency distribution of SSI wallet cryptographic functions - 100 times - 3 rounds.}
    \label{fig:moustache-ssi}
\end{figure}

Figure \ref{fig:moustache-ssi} shows the latency distribution of our SSI wallet functions. The three colors represent the performance results for the three iterations of the experiment. Some individual dots represent outliers, which are spikes in latency and account for less than 1\% of the measurements. Table \ref{tab:ssi-values} provides a statistical summary of the latencies of SSI operations based on the results of 300 runs of each function.


As the SSI wallet is implemented as a browser extension, the performance variance is likely due to the single-threaded nature of browser event loops and the overhead from the Web Crypto API. In matter of fact, most SSI wallet stacks (Aries, Veramo) are designed for mobile environments and are rarely optimized for browser execution. Moreover, background service workers experience cold starts \cite{chrome}, but improve over time thanks to the JavaScript engine applying Just-In-Time (JIT) optimizations.

\begin{table}[ht]
    \centering
    \begin{tabular}{c c c c c}
         Function & min-max & median & mean & std deviation  \\
         \hline
         CREATE\_PEER\_DID & $3.5-70.7 ms$ & $29.53 ms$ & $27.65 ms$ & $16.18 ms$ \\
         \hline
         DID\_AUTH\_REQUEST & $1.1-9 ms$ & $2.4 ms$ & $2.63 ms$ & $1.18 ms$ \\
         \hline
         GENERATE\_CONSENT\_PROOF & $2.3-15 ms$ & $5 ms$ & $5.56 ms$  & $2.61 ms$ \\
         \hline
         SIGN\_CONSENT\_DATA & $0.7-7.7 ms$ & $2.3 ms$ & $2.73 ms$ &  $1.45 m$s\\
         \hline
         VERIFY\_CONSENT\_SIGNATURE & $1.7-14.1 ms$ & $3.85 ms$ & $4.57 ms$ & $2.31 ms$\\
         \hline
    \end{tabular}
    \caption{statistical summary of the durations of SSI wallet cryptographic functions (Figure \ref{fig:moustache-ssi})
}
    \label{tab:ssi-values}
\end{table}
As shown in Figure \ref{fig:moustache-ssi} and Table \ref{tab:ssi-values}, the CREATE\_PEER\_DID operation is the most resource-intensive, with a median around $\sim 30 ms$. The standard deviation for this operation is also as high as $16 ms$, which is more than half of the median itself. This suggests that key generation is highly sensitive to environmental variables, especially computation and resource contention. In contrast, the results for the other SSI operations show more stable performance (low standard deviation). For the rest of the paper, we will use the median latency values for each SSI operation. Most SSI operations have a latency of under $5 ms$.

\subsection{Performance analysis for Hyperledger Fabric smart contracts}
\label{blockchain-perfs}

Table \ref{tab:caliper1} shows the performance results for the first two rounds, and demonstrates the robustness and performance of our blockchain-based dynamic consent model. The system maintains high throughput with a negligible failure rate: only 3 revocation requests of the total 2,000 transactions failed.
Moreover, the throughput almost matches the send rate itself, indicating that the blockchain can satisfy all requests from web portal and research organizations. Increasing the transaction send rate x10 to simulate instant 10-fold growth in requests enables us to test the dynamic consent model's scalability under stress. As shown in Table \ref{tab:caliper2}, we achieved a maximum throughput of almost 250 TPS for write operations and around 320 TPS for queries. Only 8 revocation requests failed out of 5,000 sent, with no failures in publication of consent or in the queries under any conditions. Our revocation failure rate is around 0.16\%, which is statistically insignificant and attributed to concurrent transaction conflicts, whereby multiple transactions attempt to update the state of the same consent proof on the blockchain. While all the consent proofs were successfully published, some revocation requests targeted the same consent proof due to the large number of automated requests sent within a short timeframe \cite{thakkar}.

\begin{table}[ht]
    \centering
    \begin{tabular}{c c c c c c}
         \textbf{Function} & \textbf{Successful tx} &  \textbf{Failed tx} & \textbf{max-min}&\textbf{Average latency} & \textbf{Throughput}\\
         \hline
         PUBLISH\_CONSENT\_PROOF & $500$ & $0$ & $270 ms-30 ms$ & $150 ms$ & $49.8$ TPS \\
         \hline
         REVOKE\_CONSENT & $497$ & $3$ & $260 ms-40 ms$ & $150 ms$ & 49.8 TPS \\
         \hline
         QUERY\_CONSENT\_PROOF & $1000$ & $0$ & $20 ms-0 ms$ & $10 ms$ & $100.1$ TPS  \\
         \hline
    \end{tabular}
    \caption{Blockchain functions performance results - 2 rounds - $50$ TPS for a total of 500 write operations (PUBLISH\_CONSENT\_PROOF and REVOKE\_CONSENT), and 100 TPS for a total of 1,000 read operations (QUERY\_CONSENT\_PROOF).}
    \label{tab:caliper1}
\end{table}

\begin{table}[ht]
    \centering
    \begin{tabular}{c c c c c c}
         \textbf{Function} & \textbf{Successful tx} &  \textbf{Failed tx} & \textbf{max-min}&\textbf{Average latency} & \textbf{Throughput}\\
         \hline
         PUBLISH\_CONSENT\_PROOF & $5000$ & $0$ & $170 ms-30 ms$ & $60 ms$ & $246.8 TPS$ \\
         \hline
         REVOKE\_CONSENT & $4992$ & $8$& $150 ms-20 ms$ & $60 ms$& $230.8$ TPS \\
         \hline
         QUERY\_CONSENT\_PROOF & $10000$ & $0$ & $20 ms-0 ms$ & $10 ms$ & $327.9$ TPS  \\
         \hline
    \end{tabular}
    \caption{Summary of blockchain functions performance results under extreme stress - 1 round - 500 TPS for a total of 5,000 write operations (PUBLISH\_CONSENT\_PROOF and REVOKE\_CONSENT), and 1,000 TPS for a total of 10,000 read operations (QUERY\_CONSENT\_PROOF).}
    \label{tab:caliper2}
\end{table}

\subsection{Performance analysis of the consent model}
\label{perf-consent}
Referring to the methodology in Section\ref{methods}, we evaluate the efficiency of the consent management on ClinConNet by aggregating individual component benchmarks into the end-to-end lifecycle median time. 
The consent establishment process for a research project is the aggregation of the following sequence of operations:
\textbf{\textit{ [Create private DID (SSI) --> Connect to Cloud Agent (SSI) --> Establish connection with a research organization: DID-Auth (SSI) --> Receive Consent Form (SSI) --> Verify Form Signature (SSI) --> Sign Form (SSI) --> Generate Consent Proof (SSI) --> Authenticate with web portal (SSI) --> Publish Consent Proof (Blockchain) --> Query Consent Proof (Blockchain) --> send back consent form and CPTXRF to Research organization (SSI)]}}.

Based on the performance measures obtained in Section \ref{ssi-perfs}, the total processing time for these operations is around $200 ms$. This is well below the $1,000 ms$ threshold required for an uninterrupted user experience \cite{response-time}.
Moreover, since the most costly wallet operation is the DID creation, taking around $30 ms$, the patient can anticipate this by initializing their wallet with a number of private DIDs, so they can use one of them during consent instead of generating one during the process, effectively lowering our end-to-end lifecycle median time to $170 ms$.

\section{Conclusion}
\label{conclusion}
ClinConNet is a participant-centric platform that modernizes consent management in clinical trials and e-health in general, by integrating SSI, dynamic consent and blockchain technology.
Our proof of concept (PoC) shows that high level security and privacy features, such as unlinkability, can be achieved without compromising performance or user experience. Our end-to-end lifecycle median time, which aggregates performance measures from both the SSI wallet and the blockchain smart contracts, achieved a latency of around $170-200 ms$ confirming the viability of ClinConNet's architecture for real-time clinical applications.\\

In addition to technical considerations, ClinConNet provides interesting social and economic benefits for clinical research:
\begin{itemize}
    \item Regulatory compliance: By aligning with the Right To Be Forgotten (RTBF) under the GDPR and supporting unlinkability and other privacy features, ClinConNet minimizes the legal compliance risks for researchers and research organizations. Moreover, the participant-centric nature of the platform and the SSI wallet empower participants with data ownership and portability. This native compatibility with the GDPR will encourage the use of our platform, especially in light of the eIDAS 2.0 European regulation that introduces the European Digital Identity Wallet (EUDIW). Our system is also built around a similar wallet concept, which gives us an advantage and will further strengthen the relevance of ClinConNet in the future.
    \item Economic interest: By automating the consent lifecycle using smart contracts and equipping participants with their own wallets and interfaces to manage their consent, our platform reduces administrative costs of consent management for research organizations.
    \item Boosting trust in clinical research: ClinConNet promotes a high level of trust and transparency in consent establishment and management mechanisms by eliminating data silos and giving participants control over their consent data and identity, allowing them to manage their preferences without relying on research organizations. This should improve participant recruitment and retention in research projects.
\end{itemize}

\textbf{Limitations and future work}: Although our PoC demonstrated the feasibility and core logic of our proposal, there are still limitations that need to be addressed. 
The web portal is currently a centralized component of our architecture. To enhance resilience and maximize decentralization, our aim is to decentralize the web portal logic using blockchain-based smart contracts and a new variation of Ring Signatures that will maintain privacy and security features without the need for a web portal. Work on this has already started for the next iteration of the platform.\\
Finally, we believe that the European Health Data Space (EHDS) will provide an opportunity to showcase our platform for cross-border consent management in the EU. This platform leverages the European Blockchain Services Infrastructure (EBSI) and the upcoming EU Digital Identity Wallet (EUDIW), which are currently compatible with our vision of consent management as presented through ClinConNet.

In conclusion, this initial version of ClinConNet demonstrates that it is possible and necessary to move away from the outdated, clinician-centric consent management in order to improve the ethics and regulation of clinical trials, while taking advantage of the new legal and technical frameworks that are emerging in the EU and around the world.

\section*{Data availability}
\label{code-base}
We provide the full code base of ClinConNet, along with the benchmarking scripts and results, on GitHub. We also provide an additional component called "Research Organization Single Page Application (SPA)", which is a web application that allows us to simulate the organization part, such as sending signed consent forms to participants and receiving the forms and verifying them.\\
\textbf{Web portal}: \url{https://github.com/montassar-isbored/ClinConNet-Portal}
\\
\textbf{SSI wallet Edge Agent}: \url{https://github.com/montassar-isbored/ClinConNet-SSI-wallet}
\\
\textbf{SSI Cloud Agent}: \url{https://github.com/montassar-isbored/Veramo-cloud-agent}
\\
\textbf{Research Organization SPA}: \url{https://github.com/montassar-isbored/ClinConNet-Organization-SPA}
\\
\textbf{Performance Measures}: \url{https://github.com/montassar-isbored/ClinConNet-Perfs}
\\

\section*{Funding and Acknowledgments}
This work benefited from State aid managed by the
Agence Nationale de la Recherche (ANR) under the France
2030 programme, reference ANR-22-PESN-0006 (Project
TRACIA). It is also partly supported by the Chair Values and Policies of Personal Information (VPIP), Institut Mines-Telecom,  France, and International Alliance for
Strengthening Cybersecurity and Privacy in Healthcare (CybAlliance, Project no. 337316).

\section*{AI Declaration}
Google Gemini 2.5 Pro was used to help generate the code base of ClinConNet PoC.

\bibliographystyle{unsrt}

\end{document}